\documentclass[reqno,11pt]{amsart}

\usepackage{amssymb,amsthm}
\usepackage{amsmath}
\usepackage{amsfonts}
\usepackage{dsfont}

\usepackage[a4paper,  margin=3.5cm]{geometry}

\usepackage[active]{srcltx}
\makeatletter\@addtoreset{equation}{section}\makeatother

\newtheorem{theorem}{Theorem}[section]

\newtheorem{acknowledgement}[theorem]{Acknowledgement}

\numberwithin{equation}{section}

\title[Equilibrium States,  Phase Transitions and Dynamics  in Quantum Crystals]{Equilibrium States, Phase Transitions and Dynamics in Quantum Anharmonic
Crystals}

\author{ Yuri  Kozitsky}

\address{Instytut Matematyki, Uniwersytet Marii Curie-Sk{\l}odowskiej, 20-031 Lublin, Poland}
\email{jkozi@hektor.umcs.lublin.pl}

\keywords{KMS state; path measure; stochastic process; Green
function}
\begin{document}

\subjclass{82B10; 82B20; 82B26}%

\begin{abstract}
The basic elements of the mathematical theory of states of thermal
equilibrium of infinite systems of quantum anharmonic oscillators
(quantum crystals) are outlined. The main concept of this theory is
to describe the states of finite portions of the whole system (local
states) in terms of stochastically positive KMS systems and path
measures. The global states are constructed as Gibbs path measures
satisfying the corresponding DLR equation. The multiplicity of such
measures is then treated as the existence of phase transitions. This
effect can be established by analyzing the properties of the
Matsubara functions corresponding to the global states. The
equilibrium dynamics of finite subsystems can also be described by
means of these functions. Then three basic results of this theory
are presented and discussed: (a) a sufficient condition for a phase
transition to occur at some temperature; (b) a sufficient condition
for the suppression of phase transitions at all temperatures
(quantum stabilization); (c) a statement showing how the phase
transition can affect the local equilibrium dynamics.

\end{abstract}

\maketitle

\section{Generalities}
\label{sec:1}

In recent years, remarkable progress has been made in the
experimental testing of the fundamentals of quantum physics, as well
as in developing quantum information theory and basics of quantum
computing, see \cite{BF}. Due to these advances the elaboration of
the mathematical background of quantum theory returned to the circle
of actual tasks of applied mathematics. Developing the statistical
description of infinite systems of interacting quantum particles is
one of them. Essential results in this direction were obtained by
means of methods developed in stochastic analysis, mostly in the
approach in which states of such systems are  constructed as
probability measures on infinite dimensional path spaces. A
substantial part of these results appeared due to Michael
R\"ockner's research activity, see, e.g., \cite{A0,A00,A,A1,A2,R}.
The aim of this work is to outline the main aspects of the theory of
equilibrium states of quantum anharmonic crystals obtained in
\cite{A} in the approach based on path measures.

\subsection{The Anharmonic Crystal}
An anharmonic oscillator is a mathematical model of a point particle
moving in a potential field with multiple minima and sufficient
growth at infinity. In the simplest case, the  motion is
one-dimensional and the potential has two minima  (wells) separated
by a potential barrier. If the motion is governed by the laws of
classical mechanics, the oscillator's states are characterized by a
couple $(q,p)\in \mathds{R}^2$, where $q$ is the displacement of the
oscillator from a certain point and $p$ is its momentum (amount of
motion). In the states with sufficiently small fixed  $|p|$, the
particle is confined to one of the wells. This produces a degeneracy
-- the multiplicity of states $(q,p)$ with the same $p$ and energy
$E$, that is, the multiplicity of $q$ solving the equation
\begin{equation}
  \label{1}
H (p, q) := \frac{1}{2 m} p^2 + \frac{a}{2} q^2 + V(q) = E.
\end{equation}
Here $H (p, q)$ is the particle's Hamiltonian in which $m>0$ is the
mass of the particle and the second and third terms constitute the
potential energy. If $V\equiv 0$, the oscillator is harmonic (of
rigidity $a>0$), i.e., the third term can be considered as an
anharmonic correction to the potential energy. In the quantum case,
the particle's states are vectors of unit norm belonging to the
complex Hilbert state $L^2(\mathds{R})$. The displacement and
momentum are then unbounded operators defined on $L^2(\mathds{R})$,
satisfying (on a common domain) the following commutation relation
\begin{equation}
  \label{2}
  [p, q] := p q - q p = - {\rm i}, \qquad {\rm i} := \sqrt{-1}.
\end{equation}
In (\ref{2}), we use the physical units in which the Planck constant
is set $\hslash =1$. Assume now that an infinite system of such
particles is arranged into a crystal. That is, each particle is
attached to its own crystal site $\ell\in \mathds{Z}^d$, $d\geq 1$,
and performs oscillations in its own copy of $\mathds{R}$. For a
finite $\Lambda \subset \mathds{Z}^d$, the state space of the
particles attached to the  sites in $\Lambda$ is the tensor product
of the single-particle spaces, i.e., $\mathcal{H}_\Lambda =
L^2(\mathds{R}^\Lambda)$. The Hamiltonian of this portion of
particles is
\begin{eqnarray}
  \label{3}
 H_\Lambda & = & \sum_{\ell\in \Lambda}H_\ell + J\sum_{\ell\sim \ell', \Lambda} q_\ell q_{\ell'}, \\[.2cm] \nonumber H_\ell &:= & \frac{1}{2m}p_\ell^2 +
 \frac{a}{2}q_\ell^2 + V(q_\ell) .
\end{eqnarray}
Here $H_\ell$ is the Hamiltonian of an isolated quantum anharmonic
oscillator. The sum in the second term of the first line in
(\ref{3}) is taken over all pairs of $\ell, \ell'\in \Lambda$
satisfying $|\ell - \ell'|=1$. It describes the interaction between
the neighboring oscillators located in $\Lambda$ with intensity
$J>0$. The anharmonic potential $V$ is assumed to grow at infinity
faster than $q^2$. For simplicity, in this article we take it in the
form
\begin{equation}
  \label{4}
V(q) = - b_1 q^2 + b_2 q^4, \qquad b_1 , b_2 >0.
\end{equation}
The Hamiltonian $H_\Lambda$ in (\ref{3}) with $V$ as in (\ref{4})
can be defined as a lower bounded self-adjoint operator in
$\mathcal{H}_\Lambda$ such that $\exp (-\beta H_\Lambda)$ is a
positive trace-class operator for each $\beta
>0$. Thus, one can set
\begin{equation}
  \label{5}
Z_{\beta , \Lambda}= {\rm trace} \exp \left(-\beta H_\Lambda\right).
\end{equation}
The state of thermal equilibrium of the oscillators attached to the
sites in $\Lambda$ (local Gibbs state) is defined as a positive
normalized linear functional $\varrho_{\beta , \Lambda}:
\mathfrak{C}_\Lambda \to \mathds{C}$ by the following formula
\begin{equation}
  \label{6}
\varrho_{\beta , \Lambda} (A) = {\rm trace}\left[A  \exp
\left(-\beta H_\Lambda\right)\right]\bigg{/} Z_{\beta, \Lambda},
\qquad A \in \mathfrak{C}_\Lambda.
\end{equation}
Here $\beta = 1/k_{\rm B} T$, $k_{\rm B}$ and $T$ are Boltzmann's
constant and temperature, respectively, and $\mathfrak{C}_\Lambda$
is the algebra of all bounded linear operators $A:
\mathcal{H}_\Lambda \to \mathcal{H}_\Lambda$, called observables. By
H{\o}egh-Krohn's theorem \cite[page 72]{A} $\rho_{\beta, \Lambda}$
is uniquely determined by its values on the linear span of products
\begin{equation*}
\mathfrak{a}^\Lambda_{t_1 }(F_1) \cdots \mathfrak{a}^\Lambda_{t_n
}(F_n), \qquad n\in \mathds{N}, \quad F_1 , \dots , F_n \in
\mathfrak{F}_\Lambda, \quad t_1 , \dots , t_n \in \mathds{R},
\end{equation*}
where $\mathfrak{F}_\Lambda$ is a \emph{complete} �family of
multiplication operators by bounded measurable functions $F :
\mathds{R}^\Lambda \to \mathds{C}$, whereas
\begin{equation*}
\mathfrak{a}^\Lambda_{t }(A) := \exp\left({\rm i} t H_\Lambda
\right) A \exp\left({-\rm i} t H_\Lambda \right), \qquad A\in
\mathfrak{C}_\Lambda.
\end{equation*}
According to  \cite[Theorem 1.3.6]{A}, $\mathfrak{F}_\Lambda$ is
complete if it satisfies: (a) for $F_1 , F_2 \in
\mathfrak{F}_\Lambda$, the point-wise products $F_1 F_2$ is also in
$\mathfrak{F}_\Lambda$; (b) the constant function $1$ belongs to
$\mathfrak{F}_\Lambda$; (c) for each distinct $x_\Lambda, y_\Lambda
\in \mathds{R}^\Lambda$, one finds $F\in \mathfrak{F}_\Lambda$ such
that $F(x_\Lambda) \neq F(y_\Lambda)$. Since $H_\Lambda$ is
self-adjoint, the map $A \mapsto \mathfrak{a}^\Lambda_{t }(A)$ is an
isometric automorphism of $\mathfrak{C}_\Lambda$. At  the same time,
the map $\mathds{R}\ni t \mapsto \mathfrak{a}^\Lambda_{t }(A)\in
\mathfrak{C}_\Lambda$ is the (time) evolution of the observable $A$.
The group $\{\mathfrak{a}^\Lambda_{t }\}_{ t\in \mathds{R} }$
describes the dynamics of the corresponding finite subsystem. The
mentioned above H{\o}egh-Krohn theorem implies that $\rho_{\beta,
\Lambda}$ is determined by the Green functions
\begin{equation}
  \label{9}
G_{F_1 , \dots F_n}^{\beta , \Lambda} (t_1 , \dots , t_n) :=
\rho_{\beta , \Lambda} \left[\mathfrak{a}^\Lambda_{t_1 }(F_1) \cdots
\mathfrak{a}^\Lambda_{t_n }(F_n) \right],
\end{equation}
with all choices of $F_1 , \dots , F_n \in \mathfrak{F}_\Lambda$.
Each Green function admits an analytic continuation  to the domain
\begin{equation}
  \label{10}
  \mathcal{D}_{n,\beta} := \{ (\zeta_1 , \dots , \zeta_n) \in \mathds{C}^n:
0< {\rm Im} (\zeta_1 )< \cdots < {\rm Im} (\zeta_n) < \beta  \}.
\end{equation}
Furthermore, see \cite[Theorem 1.2.32, page 78]{A}, it can further
be continuously extended to the closure
$\overline{\mathcal{D}}_{n,\beta}$ of (\ref{10}). The set
\begin{equation}
  \label{11}
 \mathcal{D}_{n,\beta}^{(0)} := \{ (\zeta_1 , \dots , \zeta_n) \in
 \overline{\mathcal{D}}_{n,\beta}: {\rm Re} (\zeta_1 )= \cdots = {\rm Re}
 (\zeta_n)=0 \}\end{equation}
has the following property: each two continuous functions $f_1, f_2
: \overline{\mathcal{D}}_{n,\beta} \to \mathds{C}$, analytic on
$\mathcal{D}_{n,\beta}$ and equal on $\mathcal{D}_{n,\beta}^{(0)}$,
are equal as functions. Then $G_{F_1 , \dots F_n}^{\beta , \Lambda}$
is uniquely determined by its restriction to (\ref{11}), that is, by
the Matsubara function
\begin{equation}
  \label{12}
\Gamma_{F_1 , \dots F_n}^{\beta , \Lambda} (\tau_1 , \dots , \tau_n)
= G_{F_1 , \dots F_n}^{\beta , \Lambda} ({\rm i}\tau_1 , \dots ,
{\rm i}\tau_n).
\end{equation}

\subsection{The Path Measures}
The main ingredient of the technique developed in \cite{A} is the
following representation, see \cite[Theorem 1.4.5]{A},
\begin{equation}
  \label{13}
\Gamma_{F_1 , \dots F_n}^{\beta , \Lambda} (\tau_1 , \dots , \tau_n)
= \int_{\Omega_{\beta,\Lambda}} F_1 (\omega_\Lambda (\tau_1))\cdots
F_n (\omega_\Lambda (\tau_n)) \mu_{\beta, \Lambda} (d
\omega_\Lambda).
\end{equation}
Here $\mu_{\beta, \Lambda}$ is a probability measure on the Banach
space $\Omega_{\beta,\Lambda}$ of `temperature loops', which is
\begin{gather*}
  \Omega_{\beta,\Lambda}=\{ \omega_\Lambda = (\omega_\ell)_{\ell \in
  \Lambda}: \omega_\ell \in \mathcal{C}_\beta\}, \quad
  \|\omega_\Lambda\|= \sup_{\ell \in
  \Lambda}\|\omega_\ell\|_{\mathcal{C}_\beta}, \\[.2cm] \nonumber
  \mathcal{C}_\beta = \{ \phi\in C([0, \beta]\to \mathds{R}):
  \phi(0) = \phi(\beta)\}, \quad \|\phi\|_{\mathcal{C}_\beta} =
  \sup_{\tau \in [0,\beta]} |\phi(\tau)|.
\end{gather*}
The measure $\mu_{\beta, \Lambda}$ is constructed in the following
way. Let
\begin{equation}
  \label{15}
H^{\rm har} = \frac{1}{2m} p^2 + \frac{a}{2} q^2
\end{equation}
be the Hamiltonian of a single harmonic oscillator, cf. (\ref{1})
and (\ref{3}), which can be defined as an unbounded self-adjoint
operator on $L^2(\mathds{R})$. It has discrete spectrum consisting
of nondegenerate eigenvalues
\begin{equation}
  \label{16}
E_n^{\rm har} = (n + 1/2) \varDelta^{\rm har}, \qquad \varDelta^{\rm
har} = \sqrt{a/m},
\end{equation}
see \cite[Proposition 1.1.37, page 41]{A}. Set $Z^{\rm har}_\beta =
{\rm trace } \exp(-\beta H^{\rm har})$, cf. (\ref{5}), and then
\begin{gather}
  \label{17}
S_\beta(\tau, \tau') = {\rm trace}\bigg{(}q e^{-|\tau - \tau'|H^{\rm
har}}q e^{-(\beta -|\tau - \tau'|)H^{\rm har}}  \bigg{)}\bigg{/}
Z_\beta^{\rm har} \\[.2cm] \nonumber = \bigg{(} e^{-|\tau - \tau'|\varDelta^{\rm
har}}+ e^{-(\beta -|\tau - \tau'|)\varDelta^{\rm har}}
\bigg{)}\bigg{/} 2 \sqrt{ma} \bigg{(}1 - e^{-\beta \varDelta^{\rm
har}} \bigg{)}, \quad \tau, \tau' \in [0,\beta].
\end{gather}
By means of the `propagator' (\ref{17}) we define a Gaussian
measure, $\chi_\beta$, on $\mathcal{C}_\beta$ by its Fourier
transform
\begin{eqnarray*}
& & \int_{\mathcal{C}_\beta} \exp\left({\rm i} \int_0^\beta f(\tau)
\phi(\tau) d \tau \right)\chi_\beta ( d \phi)\\[.2cm] \nonumber & & \quad = \exp\left(
-\frac{1}{2} \int_0^\beta \int_0^\beta S_\beta(\tau, \tau') f(\tau)
f(\tau') d \tau d \tau'\right), \quad f \in \mathcal{C}_\beta,
\end{eqnarray*}
see \cite[pages 99 and 125]{A}. Let $\chi_{\beta,\Lambda}$ be the
Gaussian measure  on $\Omega_{\beta, \Lambda}$ defined as the
product of the corresponding copies of $\chi_\beta$. Then the path
measure in (\ref{13}) is
\begin{equation}
  \label{19}
\mu_{\beta, \Lambda}( d \omega_\Lambda) = \frac{1}{N_{\beta,
\Lambda}}\exp\bigg{(}- I_{\beta , \Lambda}(\omega_\Lambda)
\bigg{)}\chi_{\beta, \Lambda}( d \omega_\Lambda),
\end{equation}
where $N_{\beta, \Lambda}$ is the normalization factor and
\begin{equation*}
I_{\beta , \Lambda}(\omega_\Lambda)= - J \sum_{\ell \sim \ell',
\Lambda} \int_0^\beta \omega_\ell (\tau) \omega_{\ell'} (\tau) d
\tau + \sum_{\ell \in \Lambda} \int_0^\beta V(\omega_\ell (\tau)) d
\tau.
\end{equation*}
Note that by (\ref{13}), (\ref{12}) and then by (\ref{9}) the
measure (\ref{19}) uniquely determines the state (\ref{6}). That is,
the local states (\ref{6}) can be constructed as Gibbs measures,
similarly as in the case of classical anharmonic crystals. Here,
however, the classical variable $q_\ell\in \mathds{R}$ is replaced
by a continuous  path $\omega_\ell$,  which is an element of an
infinite dimensional vector space,  $\mathcal{C}_\beta$. Going
further in this direction, one can define global Gibbs states  of
the quantum crystal as the probability measures on the space of
tempered configurations $\Omega^{\rm t}_\beta$ satisfying the
Dobrushin-Lanford-Ruelle (DLR) equation, see \cite[Chapter 3]{A}. It
can be shown, see \cite[Theorem 3.3.6]{A} or \cite[Theorem 3.1]{KP},
that the set of all such measures, which we denote by
$\mathcal{G}_\beta$, is a nonempty weakly compact simplex with a
nonempty extreme boundary ${\rm ex}(\mathcal{G}_\beta)$.  By virtue
of the DLR equation, the set $\mathcal{G}_\beta$ can contain either
one or infinitely many elements. Correspondingly, the multiplicity
(resp. the uniqueness) of the Gibbs states existing at a given value
of the temperature means that $|{\rm ex}(\mathcal{G}_\beta)|
> 1$ (resp. $|\mathcal{G}_\beta| = 1)$. In the physical
interpretation, the multiplicity corresponds to a phase transition,
cf. \cite[Chapter 7]{A}.

For a finite $\Lambda\subset \mathds{Z}^d$, let
$\mathfrak{M}_\Lambda$ be the subset of $\mathfrak{C}_\Lambda$
consisting of all multiplication operators by $F\in
L^\infty(\mathds{R}^\Lambda)$. Note that $\mathfrak{M}_\Lambda$ is a
maximal $C^*$-subalgebra of $\mathfrak{C}_\Lambda$. Each such an $F$
can be considered as a function $F:\mathds{R}^{\mathds{Z}^d}\to
\mathds{C}$.  Set
\begin{equation*}
  \mathfrak{M}= \bigcup_{\Lambda} \mathfrak{M}_\Lambda,
\end{equation*}
where the union is taken over all finite $\Lambda\subset
\mathds{Z}^d$. For $F_1 , \dots , F_m \in \mathfrak{M}$ and  $\mu
\in \mathcal{G}_\beta$, the Matsubara function corresponding to
these $F_i$ and $\mu$ is
\begin{equation}
  \label{22}
\Gamma^\mu_{F_1 , \dots , F_n} (\tau_1 , \dots , \tau_n) =
\int_{\Omega_\beta^{\rm t}} F_1 (\omega (\tau_1)) \cdots
F_n(\omega(\tau_n))\mu(d\omega),
\end{equation}
where $\tau_1, \dots , \tau_n \in [0,\beta]$. Then $\mu$ is said to
be $\tau$-shift invariant if, for each $\vartheta\in [0,\beta]$, the
following holds
\begin{equation}
  \label{23}
  \Gamma^\mu_{F_1 , \dots , F_n} (\tau_1 +\vartheta, \dots , \tau_n+\vartheta)
  = \Gamma^\mu_{F_1 , \dots , F_n} (\tau_1 , \dots , \tau_n),
\end{equation}
where the addition is modulo $\beta$. Let $\mathcal{G}_\beta^{\rm
phase}$ be the  subset of ${\rm ex}(\mathcal{G}_\beta)$ consisting
of all $\tau$-shift invariant measures. Its elements are called
\emph{thermodynamic phases} or \emph{states of thermal equilibrium}
of the quantum crystal. Each $\mu \in \mathcal{G}_\beta^{\rm phase}$
is defined by its Matsubara functions (\ref{22}) corresponding to
all possible choices of $n\in \mathds{ N}$ and $F_1 , \dots , F_n
\in \mathfrak{M}$, cf. \cite{Birke}. If $\mathcal{G}_\beta$ is a
singleton, then clearly $\mathcal{G}_\beta=\mathcal{G}_\beta^{\rm
phase}$. A state $\mu \in \mathcal{G}_\beta^{\rm phase}$ is called
\emph{translation invariant} if its Matsubara functions are
invariant with respect to the shifts of the lattice $\mathds{Z}^d$.

\section{The Results}

Now we present three main results concerning the properties of the
set $\mathcal{G}_\beta^{\rm phase}$.
\subsection{Phase Transitions and Quantum Stabilization}
It can be shown, see \cite[Theorem 3.7.4]{A} or \cite[Theorem
3.8]{KP}, that there exist translation invariant $\mu^{\pm} \in
\mathcal{G}_\beta^{\rm phase}$ such that, for each $\ell \in
\mathds{Z}^d$ and $\mu \in \mathcal{G}_\beta^{\rm phase}$, the
following holds
\begin{equation}
  \label{24}
M^{-} \leq M^\mu_\ell \leq M^{+}, \qquad M^{-} = - M^{+},
\end{equation}
where
\begin{equation}
  \label{25}
  M_\ell^\mu = \int_{\Omega_\beta^{\rm t}} \omega_\ell (\tau) \mu(
  d\omega),
\end{equation}
and $M^{\pm} = M_\ell^{\mu^{\pm}}$. In view of (\ref{23}), the
integral in (\ref{25}) is independent of $\tau$. By (\ref{24}) we
have that $M^{+} = M^{-} =0$ whenever  $\mathcal{G}_\beta$ is a
singleton and $M^{+} >0$ implies that $|\mathcal{G}_\beta^{\rm
phase}|>1$. Moreover, $M^{+} =0$ is also sufficient for
$|\mathcal{G}_\beta|=1$, see \cite{A0,A00}. Assume that the lattice
dimension satisfies $d\geq 3$. Set
\begin{equation*}
E(p) = \sum_{j=1}^d \left[ 1 - \cos p_j\right], \qquad \theta (d) =
\frac{d}{(2\pi)^d}\int_{(-\pi, \pi]^d} \frac{d p}{E(p)}.
\end{equation*}
It is possible to show that $\theta(d)>1$ for all $d\geq 3$ and
$\theta (d) \to 1^{+}$ as $d\to +\infty$. For $u\in [0,1)$, set
$t(u) = (\sqrt{u}/2) [\log(1+\sqrt{u}) - \log(1-\sqrt{u})]$. Then
$t$ is an increasing function and $\lim_{u\to 1^{-}} t(u) =
+\infty$. Let $u(t)$, $t\in \mathds{R}_{+}$ be its inverse, which is
an increasing function such that $\lim_{t\to +\infty}u(t) =1$. For
$b_1$, $b_2$ as in (\ref{4}) and $a$ as in (\ref{15}), set
\begin{equation*}
  \upsilon = \frac{2 b_1 -a}{12b_2}.
\end{equation*}
Note that $\upsilon> 0$ whenever $b_1 > a/2$, and thereby the
potential energy in (\ref{1}) has two wells. Recall that $J>0$ is
the intensity of the interaction of a given pair of oscillators, see
(\ref{3}). Then $\widehat{J}:=2 d J$ is the intensity of the
interaction of a given oscillator with all its neighbors. The next
statement, cf. \cite[Theorem 3.1]{KKK} or \cite[Theorem 6.3.6]{A},
gives a sufficient condition for the existing of phase transitions
in our model.
\begin{theorem}
  \label{1tm}
For $d\geq 3$, assume that $ 4 m \upsilon^2 \widehat{J}
> \theta (d)$, and hence the equation
\begin{equation}
  \label{28}
4 m \upsilon^2 \widehat{J} u\left(\beta/ 4 m \upsilon \right)=
\theta (d)
\end{equation}
has a unique solution, $\beta_*$. Then $|\mathcal{G}_\beta^{\rm
phase}|>1$ whenever $\beta
>\beta_*$.
\end{theorem}
As follows from Theorem \ref{1tm}, the absence of phase transitions,
i.e., the fact that $|\mathcal{G}_\beta^{\rm phase}|=1$ for all
$\beta >0$ implies $ 4 m \upsilon^2 \widehat{J} \leq \theta (d)$. In
order to get the corresponding sufficient condition let us turn to
the spectral properties of the Hamiltonian $H_\ell$ given in the
second line of (\ref{3}), which can be defined as a self-adjoint
lower bounded operator in $L^2(\mathds{R})$. By \cite[Proposition
4.1]{KKK} or \cite[Theorem 1.1.60]{A}, the spectrum of $H_\ell$
entirely consists of simple eigenvalues $E_n$, $n\in \mathds{N}$.
The simplicity means that each $E_n$ corresponds to exactly one
state, contrary to the classical case where the mentioned degeneracy
might occur. By means of the analytic perturbation theory for linear
operators it is possible to prove, see \cite[Theorem 4.1]{KKK}, that
$\varDelta:= \inf_{n} (E_{n+1}- E_n)$ is a continuous function of
$m\in (0,+\infty)$ such that $m^{2/3}\varDelta \to \varDelta_0$ as
$m\to 0^+$ for some $\varDelta_0>0$. Then $R_m := m \varDelta^2$ is
a continuous function of $m\in (0,+\infty)$ such that $R_m \sim
m^{-1/3}\varDelta_0^2$ as $m \to 0^{+}$. In the harmonic case
(\ref{16}), we have $R_m^{\rm har}=a$. By analogy, we call $R_m$
\emph{quantum effective rigidity}, which, however, depends on $m$ as
just discussed. The sufficient condition mentioned above is, see
\cite[Theorem 4.6]{KKK} or \cite[Theorem 7.3.1]{A}.
\begin{theorem}
  \label{2tm}
Let the parameters introduced above satisfy $\widehat{J}<R_m$. Then
$\mathcal{G}_\beta$ is a singleton for all $\beta>0$.
\end{theorem}
According to Theorem \ref{2tm}  \emph{quantum stabilization} occurs
if the interaction intensity is smaller than the effective rigidity,
see \cite{A00,A1} and Part 2 of \cite{A} for a physical
interpretation of this effect. Note that $R_m$ can be made
arbitrarily big either by making $m$ small or $\varDelta$ big (e.g.,
by making the wells closer to each other). On the other hand, it
satisfies $R_m \leq 1/4m \upsilon^2$, see \cite[Theorem 4.2]{KKK} or
\cite[Theorem 7.1.1]{A}. Therefore, $\widehat{J}< R_m$ implies that
$4m \upsilon^2 \widehat{J}<1$, cf. (\ref{28}).

\subsection{Local Dynamics}

In this subsection, we show that the dynamics of the  oscillators
indexed by the elements of a finite $\Lambda$ can be influenced by
the phase transitions described in Theorem \ref{1tm}. To this end,
we use the notion of a \emph{stochastically positive KMS system},
see \cite{KL}. Such a system is the tuple $(\mathfrak{C},
\mathfrak{B}, \{\mathfrak{a}_t\}_{t\in \mathds{R}}, \varpi)$, where
$\mathfrak{C}$ is a $C^*$-algebra; $\{\mathfrak{a}_t\}_{t\in
\mathds{R}}$ is a group of automorphisms of $\mathfrak{C}$;
$\mathfrak{B}$ is a commutative $C^*$-subalgebra of $\mathfrak{C}$
such that the algebra generated by $\cup_{t\in \mathds{R}}
\mathfrak{a}_t(\mathfrak{B})$ is $\mathfrak{C}$; $\varpi$ is a
faithful state on $\mathfrak{C}$ which is stochastically positive
and satisfies the KMS condition with some fixed $\beta>0$. The
latter means that, for each $A,B\in \mathfrak{C}$, there exists a
function, $\Phi_{A,B}(z)$, analytic in the strip $\{z\in \mathds{C}:
{\rm Im}z \in (0,\beta)\}$ and continuous on its closure, such that
$\varpi(A \mathfrak{a}_t (B)) = \Phi_{A,B}(t)$ and $\varpi(
\mathfrak{a}_t (B)A) = \Phi_{A,B}(t+{\rm i}\beta)$, holding for all
$t\in \mathds{R}$. It can be shown, cf. \cite[Theorem 2.1]{KL},
that, for each collection $A_1, \dots , A_n$ of the elements of
$\mathfrak{C}$, the Green function
\begin{equation}
  \label{29}
G^\varpi_{A_1, \dots , A_n}(t_1 , \dots , t_n) :=
\varpi(\mathfrak{a}_{t_1}(A_1) \cdots \mathfrak{a}_{t_n}(A_n)),
\qquad (t_1 , \dots , t_n)\in \mathds{R}^n,
\end{equation}
can be continued to a function analytic in the domain defined in
(\ref{10}) and continuous on its closure. The  stochastic positivity
of $\varpi$ means that, for each collection of positive elements
$F_1, \dots , F_n$ of $\mathfrak{B}$, the function defined in
(\ref{29}) satisfies
\begin{equation*}
G^\varpi_{F_1, \dots , F_n}({\rm i}\tau_1 , \dots , {\rm i}\tau_n)
\geq 0 ,\qquad 0\leq \tau_1 \leq \cdots \leq \tau_n \leq \beta.
\end{equation*}
For a finite $\Lambda \subset \mathds{Z}^d$, let us define
\begin{equation*}
  \mathfrak{D}_\Lambda = \{ Q_{u_\Lambda} = \exp\left( {\rm i}\sum_{\ell \in \Lambda} u_\ell q_\ell\right): u_\Lambda \in \mathds{Q}^\Lambda
  \},
\end{equation*}
where $\mathds{Q}$ stands for the set of rational numbers. Clearly,
$ \mathfrak{D}_\Lambda\subset \mathfrak{M}_\Lambda$ is countable and
complete. The latter follows by the fact that $
\mathfrak{D}_\Lambda$ is closed with respect to the point-wise
multiplication, contains the unit element and separates the points
of $\mathds{R}^\Lambda$. Let $\mathfrak{N}_\Lambda$ be the closure
(in the norm of $\mathfrak{M}_\Lambda$) of the set of all linear
combinations of the elements of $\mathfrak{D}_\Lambda $ with
rational coefficients. Then $\mathfrak{N}_\Lambda$ is a separable
Banach algebra. Note that $\mathfrak{N}_\Lambda$ is a proper subset
of $\mathfrak{M}_\Lambda$, dense in $\mathfrak{M}_\Lambda$ in the
$\sigma$-weak topology. For the mentioned above states $\mu^{\pm}
\in \mathcal{G}_\beta^{\rm phase}$, we have, cf. (\ref{22}), the
Matsubara functions $\Gamma^{\mu^{\pm}}_{F_1, \dots , F_n}$, $F_1 ,
\dots , F_n \in \mathfrak{Q}_\Lambda$. These functions determine two
types of dynamics of the considered portion of oscillators.
\begin{theorem}
  \label{3tm}
Let $\mathfrak{N}_\Lambda$ and $\mu^{\pm} \in \mathcal{G}_\beta^{\rm
phase}$ be as just described. Then there exist stochastically
positive KMS systems, $(\mathfrak{C}_{\pm},\mathfrak{B}_{\pm},
\{\mathfrak{a}_t^{\pm}\}_{t\in \mathds{R}}, \varpi^{\pm})$, and
injective homomorphisms, $\pi_{\pm} : \mathfrak{N}_\Lambda \to
\mathfrak{B}_{\pm}$, such that
\begin{equation}
  \label{32}
\Gamma^{\mu^{\pm}}_{F_1, \dots , F_n} (\tau_1 , \dots , \tau_n)=
G^{\varpi^{\pm}}_{\pi_{\pm}(F_1), \dots , \pi_{\pm}(F_n)} ({\rm
i}\tau_1 , \dots , {\rm i}\tau_n), \quad 0\leq \tau_1 \leq \cdots
\leq \tau_n \leq \beta,
\end{equation}
holding for all choices of $F_1, \dots , F_n\in
\mathfrak{Q}_\Lambda$.
\end{theorem}
The proof of this statement readily follows from \cite[Theorem
3.1]{Birke}, see also \cite{KD}. Its meaning can be seen from the
following fact. For $\ell_i \in \Lambda$, let
$F_{\ell_i}(q_{\ell_i})$ be real, odd and strictly positive for
$q_{\ell_i}>0$, $i=1,2,3$. Assume also that $|\mathcal{G}_\beta^{\rm
phase}|>1$, and hence $\mu^{+} \neq \mu^{-}$, see Theorem \ref{1tm}.
By the first GKS inequality, see \cite[Theorem 3.2.2]{A}, it follows
that
\begin{equation*}
\Gamma_{F_{\ell_1}, F_{\ell_2}, F_{\ell_3}}^{\mu^{+}} (\tau_1,
\tau_2, \tau_3) >0, \quad {\rm and} \quad \Gamma_{F_{\ell_1},
F_{\ell_2}, F_{\ell_3}}^{\mu^{-}} (\tau_1, \tau_2, \tau_3) <0,
\end{equation*}
for some $\tau_1, \tau_2, \tau_3$. The second inequality follows
from the first one by changing the signs of all $\omega_{\ell}$.
Then by (\ref{32}) one obtains that
\begin{equation*}
G^{\varpi^{+}}_{\pi_{+}(F_{\ell_1}), \pi_{+}(F_{\ell_2}),
\pi_{+}(F_{\ell_3})} \neq G^{\varpi^{-}}_{\pi_{-}(F_{\ell_1}),
\pi_{-}(F_{\ell_2}), \pi_{-}(F_{\ell_3})},
\end{equation*}
which means that the oscillators in $\Lambda$ distinguish between
the wells in this case, which can be experimentally detected.

\begin{acknowledgement}
The author was supported by the DFG through the SFB 701 ``Spektrale
Strukturen and Topologische Methoden in der Mathematik" and by the
European Commission under the project STREVCOMS PIRSES-2013-612669.

\end{acknowledgement}

\end{document}